\documentclass{vgtc}                          

\ifpdf
  \pdfoutput=1\relax                   
  \usepackage{graphicx}                
  \DeclareGraphicsExtensions{.pdf,.png,.jpg,.jpeg} 
\else
  \usepackage{graphicx}                
  \DeclareGraphicsExtensions{.eps}     
\fi%

\graphicspath{{figures/}{pictures/}{images/}{./}} 

\usepackage{microtype}                 
\PassOptionsToPackage{warn}{textcomp}  
\usepackage{textcomp}                  
\usepackage{mathptmx}                  
\usepackage{times}                     
\usepackage{cite}                      
\usepackage{tabu}                      
\usepackage{booktabs}                  
\usepackage{balance}
\usepackage[table, dvipsnames]{xcolor} 

\usepackage{tabularx}
\usepackage{float}
\usepackage{subfigure}

\usepackage{todonotes}
\usepackage{balance}       

\usepackage{enumitem}

\usepackage{soul}

\usepackage[most]{tcolorbox}
\newtcolorbox{highlighted}{colback=yellow,breakable}

\onlineid{8820}

\vgtccategory{Research}

\vgtcinsertpkg

\title{How Do We Measure Trust in Visual Data Communication?}

\author{Hamza Elhamdadi\thanks{e-mail: helhamdadi@umass.edu}\\ %
        \scriptsize UMass Amherst %
\and Aimen Gaba\thanks{e-mail: agaba@umass.edu}\\ %
     \scriptsize UMass Amherst
\and Yea-Seul Kim\thanks{e-mail: yeaseul.kim@cs.wisc.edu}\\ %
     \scriptsize University of Wisconsin-Madison
\and Cindy Xiong\thanks{e-mail: cindy.xiong@cs.umass.edu}\\ %
     \scriptsize UMass Amherst}

\abstract{
Trust is fundamental to effective visual data communication between the visualization designer and the reader. 
Although personal experience and preference influence readers' trust in visualizations, visualization designers can leverage design techniques to create visualizations that evoke a ``calibrated trust," at which readers arrive after critically evaluating the information presented.  
To systematically understand what drives readers to engage in ``calibrated trust," we must first equip ourselves with reliable and valid methods for measuring trust.
Computer science and data visualization researchers have not yet reached a consensus on a trust definition or metric, which are essential to building a comprehensive trust model in human-data interaction.
On the other hand, social scientists and behavioral economists have developed and perfected metrics that can measure generalized and interpersonal trust, which the visualization community can reference, modify, and adapt for our needs.
In this paper, we gather existing methods for evaluating trust from other disciplines and discuss how we might use them to measure, define, and model trust in data visualization research.
Specifically, we discuss quantitative surveys from social sciences, trust games from behavioral economics, measuring trust through measuring belief updating, and measuring trust through perceptual methods. 
We assess the potential issues with these methods and consider how we can systematically apply them to visualization research.
} 

\CCScatlist{
  \CCScatTwelve{Human-centered computing}{Visu\-al\-iza\-tion}{Visualization design and evaluation methods}{};
}

\begin{document}


\firstsection{Introduction}
\label{intro}

\maketitle

Effective visual data communication is a dynamic exchange between a visualization (and its designers) and the information consumer.
When designers create visualizations for communication, they make choices about encoding and design that they think will accurately and persuasively communicate their interpretation of the data. 
When information consumers interpret visualizations, they choose which patterns to focus on and what stories to take away.
Visual data communication involves critical thinking and trust from both parties \cite{dork2013critical}, making trust especially important to establish \cite{kelton2008trust}.
Human readers can learn to \textbf{trust} or distrust the conveyed information, and visualizations are judged more or less \textbf{trustworthy} depending on their design and delivery.

Ideally, we want human readers to engage in ``calibrated trust'' when interacting with data visualizations, which involves \textit{critically evaluating the information, rather than unconditionally dismissing or accepting it}.
When the reader does not blindly reject a visualization, they are exhibiting some trust that the visualization is worth at least contemplating.
Similarly, when the reader does not blindly accept the conclusions of the visualization, they are exhibiting the capability to detect signs of mistrust that may be present in the visualization. 
This calibrated trust exists between a visualization reader and a visualization (human-to-visualization trust) but can also be extended as trust between the visualization reader and the visualization creator (human-to-human trust).
For the purposes of this paper, we will cover only the human-to-visualization trust.

Social science research has demonstrated that by educating people to identify signs of trust, they can better calibrate themselves in identifying signs of mistrust \cite{bockler2019theory}.
As visualization researchers, we can investigate visualization components that elicit trust and mistrust to empower readers to identify signs of mistrust and guide designers to create trustworthy visualizations; thus, we optimize the effectiveness of visual data communication and data-driven decision making.

However, personal experience and preference can heavily influence trust \cite{goudge2005can}.
For example, subjectively perceived transparency \cite{xiong2019examining} and aesthetic preferences \cite{alter2009uniting} can impact perceived trustworthiness. 
Trust is also very contextualized.
Depending on the domain and application, the role of trust and peoples' criteria for trustworthiness might change \cite{kramer1999trust, rousseau1998not}.
Therefore, it is unlikely that we will easily find a general set of rules to optimize trust across all scenarios of visual data communication.
Instead, we may take a domain- or task-specific approach to look at the role of trust in different contexts.
This bottom-up approach will allow us to model trust for visualization research by understanding what visual data communication trust rules generalize across domains and tasks.

\begin{figure}[h!]
    \centering
    \includegraphics[trim={0 0 0 0},clip,width = \columnwidth]{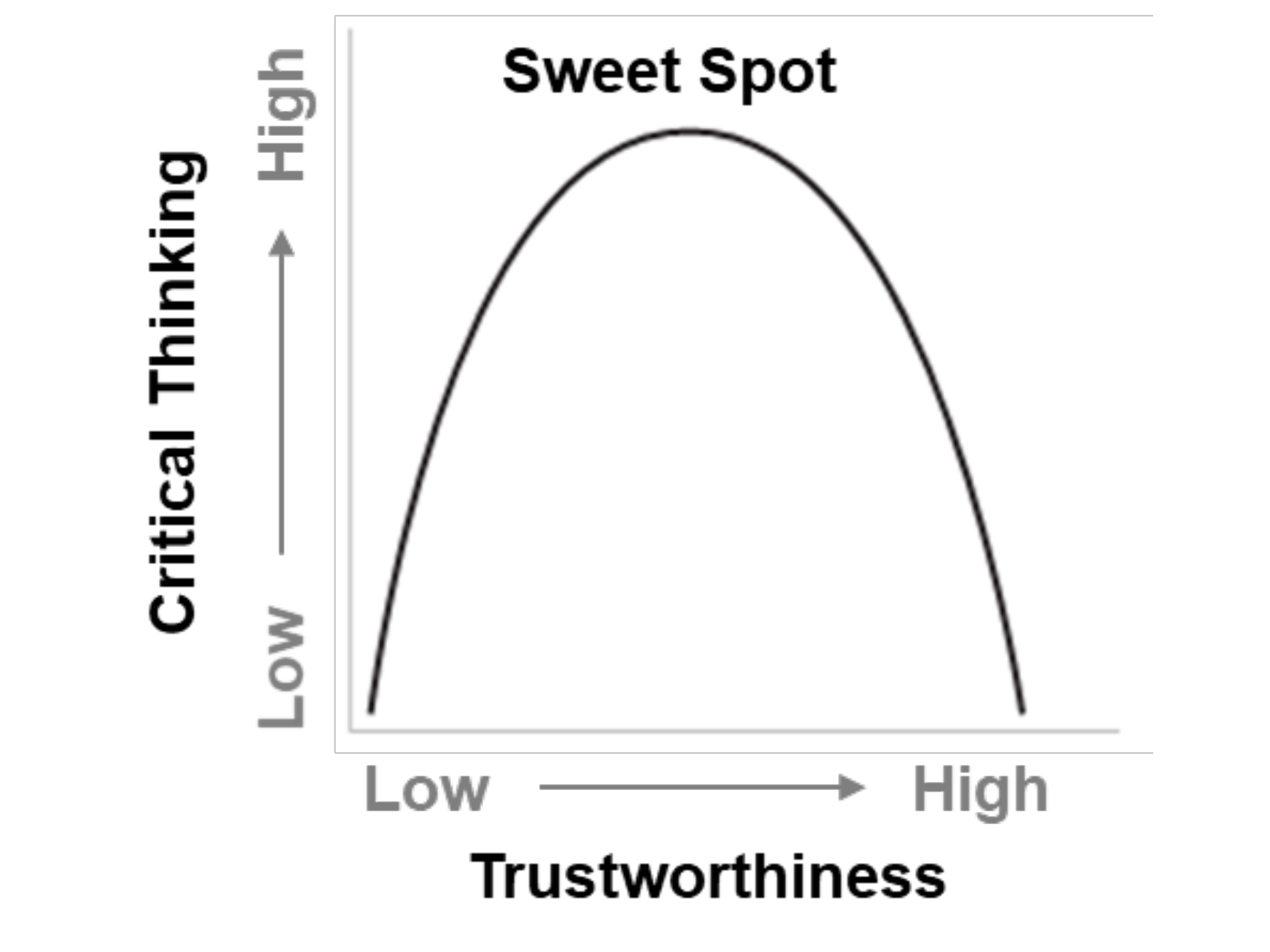}
    \caption{Extremely low and overwhelmingly high trust in visualized data might promote low critical thinking. Visualization authors should consider the trade-offs and aim for the sweet stop to maximize trust and enhance critical thinking.}
    \label{fig:TrustCurvilinear}
\end{figure}

To do so, we need to identify reliable and valid methods for measuring trust in visual data communication across different contexts and account for individual differences.
Measuring trust with nuance and accuracy requires a multi-faceted approach. 
In this paper, we survey trust measurements used in existing research from social sciences, computer sciences, and behavioral economics, reflect on how these disciplines define and measure trust, and propose ways to use these metrics in visual data communication. 
This list is by no means exhaustive, but we hope that this collection will help the visualization research community better understand the complexity of trust and the need to measure it via a multi-faceted approach. 




\section{Trust Measurement in Computer Science}

Due to the complicated nature of trust, existing computer science research varies in the approach to measuring trust, and the definition of trust varies from paper to paper \cite{artz2007survey}. 
For example, Boyd et al. define trust based on whether people believe what they see \cite{boyd2021can}.
Some define trust in terms of the entity relaying the information (i.e., is the source of this information actually who they claim to be?) \cite{artz2007survey}.
Others define trust as ``believing others in the absence of clear-cut reasons to disbelieve" \cite{rotter1980interpersonal}.

Computer scientists have measured trust with multiple approaches.
Jian et al. \cite{jian2000foundations} developed a 12-item 7-point Likert scale to measure user trust in a movie recommendation app. 
To measure people's trust in a machine learning model, Yin et al. \cite{yin2019understanding} used two metrics: how often the users agreed with the model's prediction and how frequently the users changed their prediction to be consistent with the model prediction.
To measure the impact of computer-mediated communication on interpersonal trust, Zheng et al. \cite{zheng2002trust} used a trust game referred to as the ``Day-Trader Investment Task." 
Each participant is given a fixed amount of money in a hypothetical scenario and can choose to invest a portion.
In this set-up, the amount of money a participant invests is a proxy for the amount of interpersonal trust they exhibit.

However, researchers rarely test these trust metrics for reliability and validity.
Psychologists often rigorously test their surveys to ensure the repeatability of their results and minimize harmful biases.
We highlight a case study illustrating why this process is critical for measuring trust in computer science.
In an experimental study regarding trust between humans and automated systems, Jian et al. developed a 12-item Likert scale (from 12 identified trust factors, six positive, six negative) for measuring human-computer trust \cite{jian2000foundations}. Although Jian et al. noted that testing validation was necessary, the scale was used in 179 papers (in its original form in 100) as of 2019. Despite the proliferation of this scale, Gutzwiller et al. \cite{gutzwiller2019positive} ran an experimental study to test the scale's efficacy and found that the scale biased users' trust scores depending on the ordering of questions; if the positive items were given first, the trust scores were significantly higher than if the negative items were given first or if the items were given in a random order.
This case study demonstrates that a lack of iteration on the trust metrics we use may lead to biased results; hence, a non-unified approach to measuring trust could hide a lack of repeatability or other unknown flaws with various existing trust measures.

In this paper, we propose multiple methods to measure trust grounded in theories of social sciences, behavioral economics, and cognitive sciences.
But like the example illustrated, although these methods have been tested and perfected over time in these other communities as reliable and valid measures of trust, we must iterate and improve on them to ensure their reliability and validity in the context of measuring trust in computer science research. 
We also focus on quantitative trust measures in this paper; however, this does not mean we should neglect qualitative measures.
Qualitative metrics, such as grounded theory \cite{kozinets2019netnography}, interviews, and open-text survey questions can reveal nuances and additional insights beyond quantitative methods. 
For example, Hong et al. measured trust via the qualitative coding of existing research papers in machine learning that mention trust \cite{hong2020human}. Effectively, this approach considers the trust definition used by scientists working in machine learning.
Similarly, Hu et al. measured users' trust in the output of an assembly code synthesizer via free-response questions that the users were encouraged to answer; Hut et al. evaluated user answers (e.g., ``I'm involved enough in the process that I trust the results") qualitatively to assess general trust patterns \cite{hu2021assuage}.

\begin{table*}[t!]
    \centering
    \begin{tabular}{|l|l|l|}
        \hline
        \textbf{Original Evans and Revelle Item \cite{evans2008survey}}                & \textbf{New Visualization-Appropriate Item}            \\ \hline
        Anticipate the needs of others                          & Provides useful information                            \\ \hline
        Have always been completely fair to others              & Depicts balanced data                                  \\ \hline
        Stick to the rules                                      & Follows best design practices                          \\ \hline
        Value cooperation over competition                      & Uses unbiased data sources                             \\ \hline
        Return extra change when a cashier makes a mistake      & Issues notices and revises when mistakes are found     \\ \hline
        Would never cheat on my taxes                           & Does not falsify data                                  \\ \hline
        Finish what I start                                     & Provides complete data                                 \\ \hline
    \end{tabular}
    \vspace{2mm}
    \caption{Possible modifications for Evans and Revelle survey items for visualization purposes}
    \label{tab:revelle_modifications}
\end{table*}

\section{Trust Measurement in Visualization Research}

Visualization researchers recognize the complexity of trust and its measurement and leverage various metrics to measure trust. 
The most prominent being a Likert scale - e.g., ``on a scale from 1 to 7, how much do you trust this visualization?'' \cite{xiong2019examining}, or ``on a scale of 1 to 100, how much do you trust that the base data is accurate? ''\cite{kim2017data}, or ``on a scale from 1 to 9, how much do you trust the visualized model predictions?" \cite{zhou2019effects}.
The Likert scale can vary in the number of discrete values offered (e.g., 0 to 100 \cite{kim2017data}, 1 to 5 \cite{zehrung2021vis}, or 1 to 7 \cite{jian2000foundations}), or the number of the Likert scale items (e.g., 3 items \cite{kong2019trust} or 5 items \cite{xiong2019examining}). 

Other approaches include using substitutions variables as a proxy of trust.
For example, Xiong et al. asked participants to choose the best map visualization for a firetruck in a hypothetical emergency.
They measured trust through the decision (i.e., participants likely trusted the chosen map more than the others), as well as ratings of perceived trust and transparency, which have four dimensions: the accuracy of the information, the clarity of the communicated information, the amount of information disclosed, and the extent to which the shared information is a thorough representation of the underlying data \cite{xiong2019examining}.
In \cite{kong2019trust}, the researchers measured participants' trust in title-misaligned visualizations via the three proxies: perceived credibility, perceived bias, and appropriateness.
In another example, \cite{zehrung2021vis} measured trust in human-generated versus algorithm-generated visualizations via perceived usefulness and preferences by asking participants to choose between different human-generated and algorithm-generated visualizations for a particular task. 




\subsection{Issues with Trust Measurement in Visualization}

We discuss two issues with the existing trust metrics in visualization research: lack of objectivity and lack of reliability and validity. 

The Likert scales used to measure trust are usually participants' self-reported ratings, which, like all self-report measures, are not always accurately reported nor representative of the participants' true inner state \cite{cozby2012methods}. 
Furthermore, participants' interpretations of the Likert values may differ, sometimes causing two participants exhibiting the same level of trust to choose different values on the scale. 
This effect can compound when participants cannot opt-out \cite{chita2021can} (e.g., via a N/A option).
We can mitigate these shortcomings by supplementing the quantitative measures with a qualitative open-text response box so participants can further explain the rationale behind their ratings. However, low participant motivation to detail their inner thoughts and inability to articulate their mental representation of trust can stymie the effectiveness of these qualitative responses.

We must recognize that reliable and valid measures typically result from iterations of testing, experimentation, and revision \cite{shrout2012psychometrics, wall2022vishikers}.
The current state, where trust is decomposed and defined by different researchers as different variables, it becomes difficult to evaluate whether the decomposition (and the subsequent measures) are valid, that they are actually measuring trust, or reliable, that it will yield consistent results across individuals sharing similar characteristics or within the same individual across time. 
For example, in \cite{xiong2019examining}, where both participants rated the perceived trustworthiness of maps and selected a map to follow, the researchers found inconsistent results from the two measures. 
However, Maps highly rated for perceived trust did not always coincide with the selected map, leaving questions about how these measures capture different trust components. 

Furthermore, the different designs of the Likert scale, such as including different number of questions/items between studies  (e.g., one \cite{kim2017data}, six \cite{liao2022user}, twelve \cite{jian2000foundations}), different ranges of values (e.g., 0 to 100 \cite{kim2017data}, 1 to 5 \cite{zehrung2021vis}, or 1 to 7 \cite{jian2000foundations}), can make it difficult for researchers to iterate on one another and compare experimental effect sizes to improve the reliability of these measures.
A single-item Likert scale approach may not be a reliable indication of a person's trust; ideally, the scale used would have more than one item, approaching the multi-faceted trust question from multiple angles to increase reliability and validity \cite{evans2008survey}. 
Additionally, while researchers frequently use scales with a higher number of items, the reasoning for choosing a particular scale and its efficacy are not often comprehensive. One of the most common (12-item) scales used \cite{kim2017data} was shown to bias the responses of participants \cite{gutzwiller2019positive}.
Visualization research should evaluate the existing n-item Likert scales and increase the reliability of the most promising scales via further iterative reliability testing on trust, changing the number of values, the number of questions, and even the wording of the questions when necessary.



One factor that might have contributed to the validity issue of trust measurement is that trust has been ill-defined in visual data communication and computer science research.
For example, trust has been defined as strongly related to bias and credibility \cite{kong2019trust}, integrity and deception and reliability \cite{jian2000foundations}), or transparency \cite{xiong2019examining}.
But studies where participants rate trust directly (e.g., ``on a scale of 1 to 100, how much do you trust that the base data is accurate" \cite{kim2017data}), participants are usually not provided a clear definition of trust, and thus it remains unclear what kind of trust they are rating if it is trusted at all.

However, we recognize this is not a unique concern to the visualization research community.
Social scientists, behavioral economists, and psychologists have operationalized trust differently \cite{rousseau1998not}.
Trust definitions include a ``willingness to be vulnerable to exploitation within a social interaction" \cite{levine2018trustworthy}, reciprocity (i.e., willingness to give up something in return) \cite{zheng2002trust}, and reliance (e.g., on a machine learning model's recommendation \cite{hong2020human}.

Social and political research defines trust via top-down and bottom-up theories. Bottom-up theories propose that socio-political trust stems from individuals that initially trust until trauma or difficult decisions cause them to reevaluate. Top-down theories argue that trust results from societal or political factors like good government, general wealth, or happiness.
Further theorizing has determined that social and political trust are also context-dependent (i.e., individuals evaluate trust in others on a case-by-case basis); for example, someone may trust their friend to take care of their pet but not to water their garden for a month \cite{uslaner2018oxford}.

\section{Measuring Trust in Social Sciences}
\label{section:soc}

Much research has been done in the social sciences to define trust and create a robust trust metric.
Trust has been measured as ``Generalized Trust'' in other people in the World Value Survey, which asks, via an open-text response box, `Generally speaking, would you say that most people can be trusted or that you need to be very careful in dealing with people?'' 
The responses are coded into two categories: those with high Generalized Trust who say that most people can be trusted, and those with low Generalized Trust who say one needs to be very careful in dealing with people \cite{haerpfer2020world}.
The responses from the Generalized Trust survey questions positively correlate with other, more objective measures of trust, such as an ``Investment Game" scenario (see Section \ref{trustgames}) \cite{johnson2012much}.

Although much of the research on trust in the social sciences are concerned with trusting behavior and general perspectives on trust and trustworthiness, similar studies have been aimed at capturing trust in a particular individual.
Everett et al. measured how trusting people are of a particular leader using a 2-item 7-point Likert scale (``How trustworthy do you think this person is?" and ``How likely would you be to trust this person's advice on other issues?"), and a voting task, where participants were asked to cast a vote between two possible leaders who had the opportunity to embezzle money from a donation to a charity \cite{everett2021moral}. 
However, trust was not explicitly defined in this survey. 
A more well-defined trust question in a different context could be ``How much do you trust your physician to perform necessary medical tests and procedures regardless of cost?" from \cite{goudge2005can}, as this question asks about a particular scenario (performing medical tests/procedures) rather than general trust.

Psychologists have argued that these single-item or two-item scales may be insufficient to reliably capture trust \cite{evans2008survey}.
Evans and Revelle \cite{evans2008survey}, for example, define trust as an enduring trait in people (rather than a transient one) that comprises the intention to accept vulnerability based upon the positive expectations of the intentions or behavior of another'' \cite{rousseau1998not}.
They argue that because trust is an enduring trait, we should assess trust in similar ways as personality and study trust as a broad construct that complements personality models such as the Big Five \cite{rammstedt2007measuring, digman1997higher}.
The Big Five Inventory is a 44-item scale that measures five dimensions of personality (extraversion, agreeableness, openness, conscientiousness, and neuroticism); each item describes a trait using different adjectives concerning a prototypical behavior, and the participant can rate via a 5-point scale indicating whether they disagree strongly (1) or agree strongly (5).
For example, one item may state ``I persevere until the task is finished''\cite{john1999big}.  
Based on this, Evans and Revelle created (and tested) a 21-item scale that measures interpersonal trust \cite{evans2008survey}.
Sample items on the scale include ``Listen to my conscience" and ``Avoid contact with others."

Some other trust surveys include the one by Justwan et al., which uses the general trust question responses and 19 variables that correlate with social/political trust (e.g., political rights, corruption, income inequality, fertility rate) as factors in a Bayesian model, which measures trust in the context of international relations \cite{justwan2018measuring}.
The World Values Survey has also made updates to now contain a set of questions (answered via a 4-point Likert scale) regarding trust/confidence in specific social groups (e.g., family, neighborhood) and organizations (e.g., churches, the press, television) \cite{haerpfer2020world} that provide more narrow focus on what elements of trust may be affecting generalized trust.

\subsection{How we can do this in visualization research}

We propose that visualization researchers can also modify interpersonal trust surveys, such as the one by Evans and Revelle \cite{evans2008survey}, to be about trust in human-data interaction.
For example, the survey item ``[this person] stick[s] to the rules'' measures trustworthiness and can be modified as ``[this visualization] follows best design practices'' to assess trust in human interaction with visualizations. See Table \ref{tab:revelle_modifications} for a more comprehensive list of modifications to the Evans and Revelle scale (Note: many items from the Revelle scale have no visualization counterpart).

For trust measures used by social scientists that measures general trust, while it may seem irrelevant in determining a reader's trust in a visualization, generalized trust can offer a baseline for measuring trust in visual data communication.
In other words, we can compare the generalized trust of the reader to their trust behavior regarding a visualization.
A generally more trusting person may rate a visualization as more trustworthy than a generally less trusting person. But, we can approximate the trustworthiness of a visualization by computing the difference - how much trust does the visualization elicit above and beyond one's baseline level of trust?

We also recognize that trust is context-dependent.
Previously, we demonstrated that adding more context to a trust question more effectively defines trust and helps participants better understand what they are rating \cite{evans2008survey}.
\cite{goudge2005can} proposed to measure trust between physicians and patients by asking ``How much do you trust your physician to perform necessary medical tests and procedures regardless of cost?".
We can modify questions like this to be about visualizations.
For example, ``How trustworthy do you think the author of this visualization is?" or ``How likely are you to rely on the information presented in this visualization to make future decisions?"

Although we proposed multiple ways to evaluate trust in visual data communication referencing survey items from social sciences, we emphasize that future researchers who wish to measure trust using similar scales should empirically test, iterate, and improve upon them.
The wording of each question might have a biasing effect on the survey taker.
Moreover, if responses to these questions use a Likert scale, we should consider that, if given the option to select ``Not Applicable," people will opt for this answer in some scenarios \cite{chita2021can}. Failing to provide this option forces the participant to answer the question, and when the subject feels the question does not apply, the answer may not be indicative of what they believe, which could skew the results.

\begin{figure}[t!]
    \centering
    \includegraphics[width = \columnwidth]{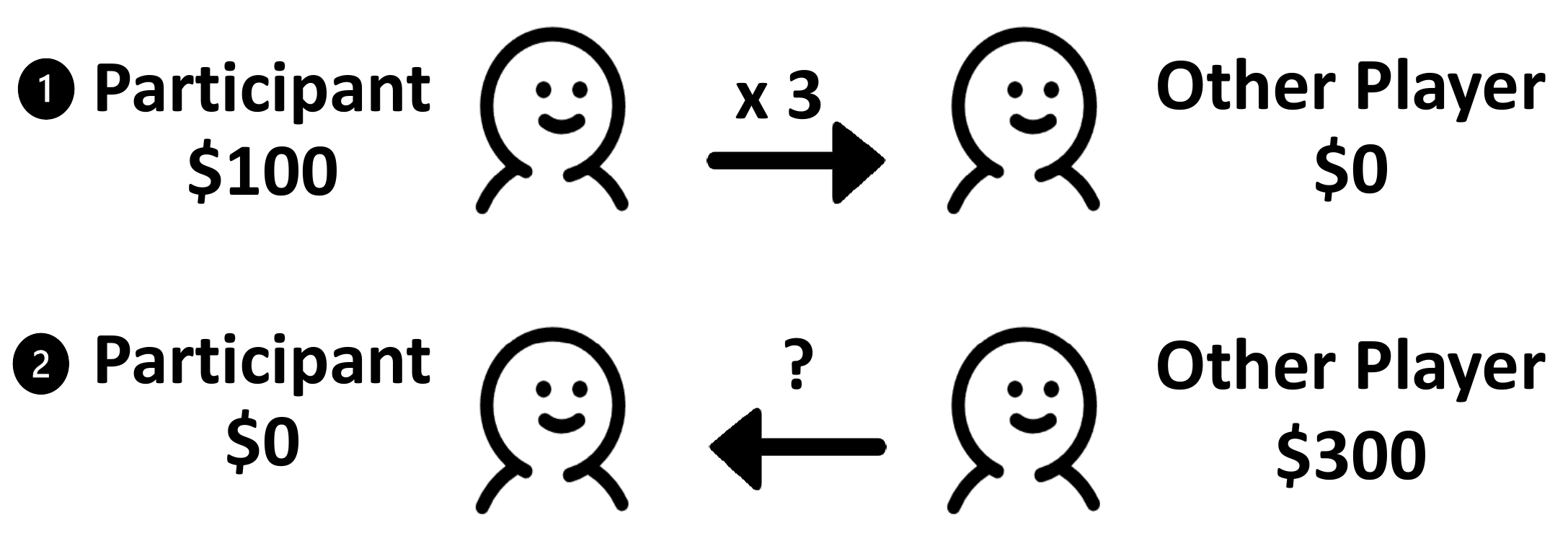}
    \caption{Trust game set-up in two steps. Top: The participant starts with some money and can decide on an amount to send to another player. The money sent gets tripled when it reaches the other player, and the other player can choose to return any amount back to the participant. The tripling is critical here because it is the lowest possible manipulation that enables the other player to share an amount of money that can profit for both players.}
    \label{fig:trustgame}
\end{figure}

\section{Trust Games in Behavioral Economics}
\label{trustgames}




Although there is no clear cross-discipline definition of trust, most behavioral economists \cite{glaeser2000measuring, zurn2017trust} adopt James S. Coleman's definition from ``Foundations of Social Theory'' \cite{coleman1990foundations}. According to this definition, trust occurs when resources are given by a trustor to a trustee with no enforceable commitment from the trustee. For example, a trustor can decide to lend their car to a trustee, knowing the risk that the trustee may not give it back. 

A subgroup of economists, particularly neoclassical economists, argue that people always seek to maximize their monetary profits and personal satisfaction without considering how their choices might affect others \cite{GOODLAND198719}. According to this view of rationality, trusting others (i.e., not thinking of maximizing one's profits) is irrational because the risks involved usually work against the person's material interests. Following this assumption, there is no reason to take the ``irrational'' risk of trusting someone else, and even simple exchanges in the real world (e.g., ordering clothes online) would require some form of complex litigation.
However, from our real-world experiences, we know this is not the case. We trust others and hope for others' trust in us.

To understand what motivates trust, behavioral economics research has employed trust games. As shown in Figure \ref{fig:trustgame}, a typical trust game involves two anonymously paired participants; one participant (the trustor) is given an amount of money from which they can choose to give a portion or none to the other participant (the trustee); the experimenter triples the trustee's money, and the trustee can give some portion or none of the resulting amount back to the trustor \cite{BERG1995122}. Note, a researcher can conduct a trust game with one participant, where the trustee is played by a computer program \cite{ANDERHUB2002197}. 
By giving money to the trustee, the trustor is willing to put themselves in a vulnerable position; hence, the implication is that a high amount of money indicates high trust.
Another possible implication is that the trustor is being altruistic and, thus, not engaging in trust. However, Brülhart and Usunier rejected altruism as the cause of ``trust-like" decisions by  \cite{brulhart2012does}.


As measuring trust via interactive games has led to multiple versions of the trust game, we discuss some of the prominent/relevant ones here.
Kreps et al. introduced a simpler version of the trust game by providing participants with only two options: to trust or not to trust \cite{kreps1990corporate}.
When a trustor decides to trust, the trustee has the option to either honor it leading to equal payoffs of $\$10$ or dishonor it leading to a payoff of $\$15$ for the trustee and -$\$5$ for the trustor. 
If the trustor decides not to trust, both the trustor and the trustee are left with $\$0$.

Guth et al. (1982) introduced a ``take-it-or-leave-it'' variation of the game, often referred to as \emph{The Ultimatum Game}.
It consists of two players given an endowment they must divide amongst themselves. 
Player One has the option to offer a part of the endowment to Player Two, which Player Two can either accept or reject. Acceptance of the proposal meant enforcement of the proposal, and rejection meant neither of the two received anything. 
The results showed that Player One made substantial offers, and Player Two rejected small but positive offers. 
Similar results occurred in a ``take-it'' variation of this game called \emph{The Dictator Game} \cite{FORSYTHE1994347}. 
In this case, Player One chooses how to allocate the endowment and Player Two could benefit from the allocation but had no power to accept or reject the decision. 
The findings showed Player One (the dictator) granted surprisingly positive amounts to the other player.
There is also \emph{the game of trust} \cite{GUTH199715} in which the first mover can decide between non-cooperation and cooperation, whereas the second mover, who decides only when cooperating, must choose whether to share the fruits of cooperation equally or not leave anything to the first mover who put the trust by cooperating in the first place.

Zheng et al. explore how trust varies by playing an interactive game called \emph{The Day-Trader Investment Task} \cite{zheng2002trust}.
The game is played in pairs in multiple trials, and the participants took part in a variant of a Prisoner’s Dilemma task, which has a history of testing group cooperation and trust \cite{komorita2019social}. 
Each participant received $\$40$ per day and imagined being a day-trader during a multi-day investment period. 
Out of that amount, they could invest all or some of it. 
The day-trader had two choices for investing the money: invest in a common pool whose payoff depended on how much the other partner invested in it, or keep it in an individual account. 
In order to conceal what each partner exactly contributed, the researcher added a random factor between -10 and +10 to represent stock market fluctuations. 
After including the random factor, each participant's investment with the group was doubled and split evenly between the two participants. 
At the end of the `day', the participants had the amount they chose to keep in their individual account, or the amount they gained from the common pool after the payout. 
The authors observed higher trust (i.e., more cooperation) in face-to-face pairs compared with pairs that never met one another.





\subsection{How we can do this in visualization research}


In \cite{zurn2017trust}, the researchers compared the trustworthiness of easy- versus difficult-to-articulate names using the trust game set-up.
Participants uniformly played with computer programs disguised as real people whose usernames varied in pronounceability.
The amount of money the participants invested represented the level of trust, and the researchers found that participants tend to trust 'other players' with easier-to-pronounce usernames.

To our knowledge, visualization research has not used trust games to measure trust, despite their providing a metric less influenced by subjective trust cognition. 
Here are some methods that visualization researchers can use to leverage trust games in their research.
We can modify the trust game to replace the second player with a visualization. Then, we can manipulate different design elements in a visualization and compare how the amount of trust changes depending on these design choices.
This set-up parallels existing findings from behavioral economics where researchers find aesthetics (or superficial) features of the other player can impact trust and will likely enable us to observe similar effects of visualization design on trust.
For example, the trust game was able to demonstrate that perceived attractiveness \cite{wilson2006judging}, age \cite{sutter2007trust}, and the width of the face \cite{stirrat2010valid} impact trust.
Similarly, we can expect the trust game to be able to identify the effect various visualization design have on perceived trust.

One issue to resolve between the traditional set-up of having two players (despite one being a computer program) and the proposed set-up, which involves only one player and visualization, is the \textit{why} the participant player should invest money into a visualization. 
Alternatively, if the game lasts multiple rounds, we will need some method to return the participant player their investment in a justifiable way, to create interaction between the player and the visualization, and potentially build (or diminish) trust.

We propose two scenarios to make the trust game set-up between a human player and a visualization more realistic.
In one scenario, the participant can make a decision on how much money to invest based on the information presented in the visualization. In this case, the more they trust the visualization, the more money they will invest.
In the second scenario, the participant can bet on whether the visualization will lead another hypothetical viewer to make a decision. If they bet correctly, they will receive some returns on their money. If not, they lose the money.
This approach tests to what extent the participant sees the visualization as trustworthy to others.
From related work on the theory of mind \cite{xiong2019curse}, we can infer that the participant's prediction of a visualization's general trustworthiness reflects how trustworthy they themselves perceive the visualization to be. 
Future work can further test these two scenarios, so the story can be iterated and improved to be comparable, reliable and valid as the traditional trust game set-up in behavioral economics research.

\begin{figure}
    \centering
    \includegraphics[trim={3cm 3cm 0 0}, clip, width=0.45\linewidth]{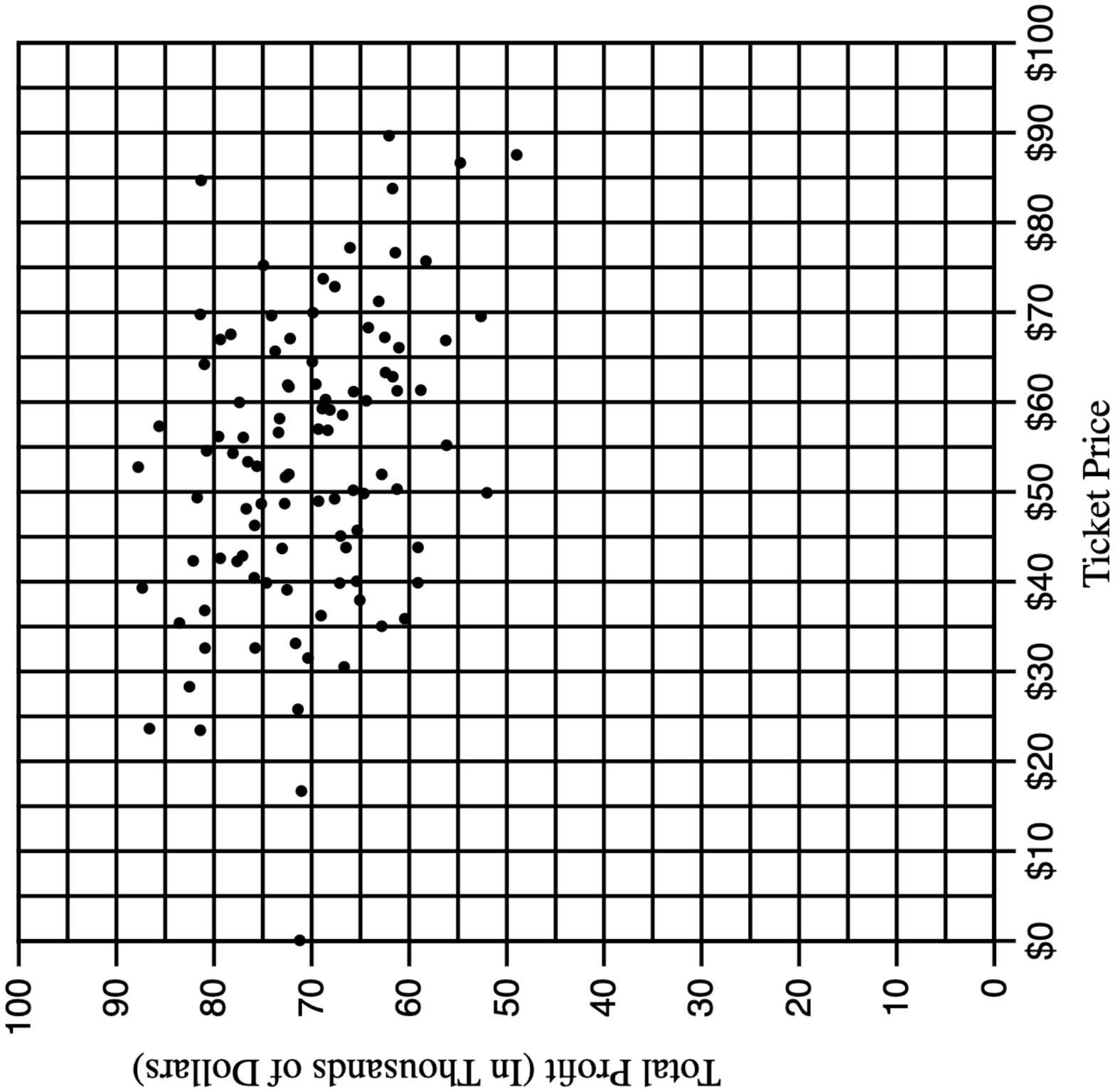}
    \includegraphics[trim={1.5cm 1.5cm 0 0}, clip, width=0.45\linewidth]{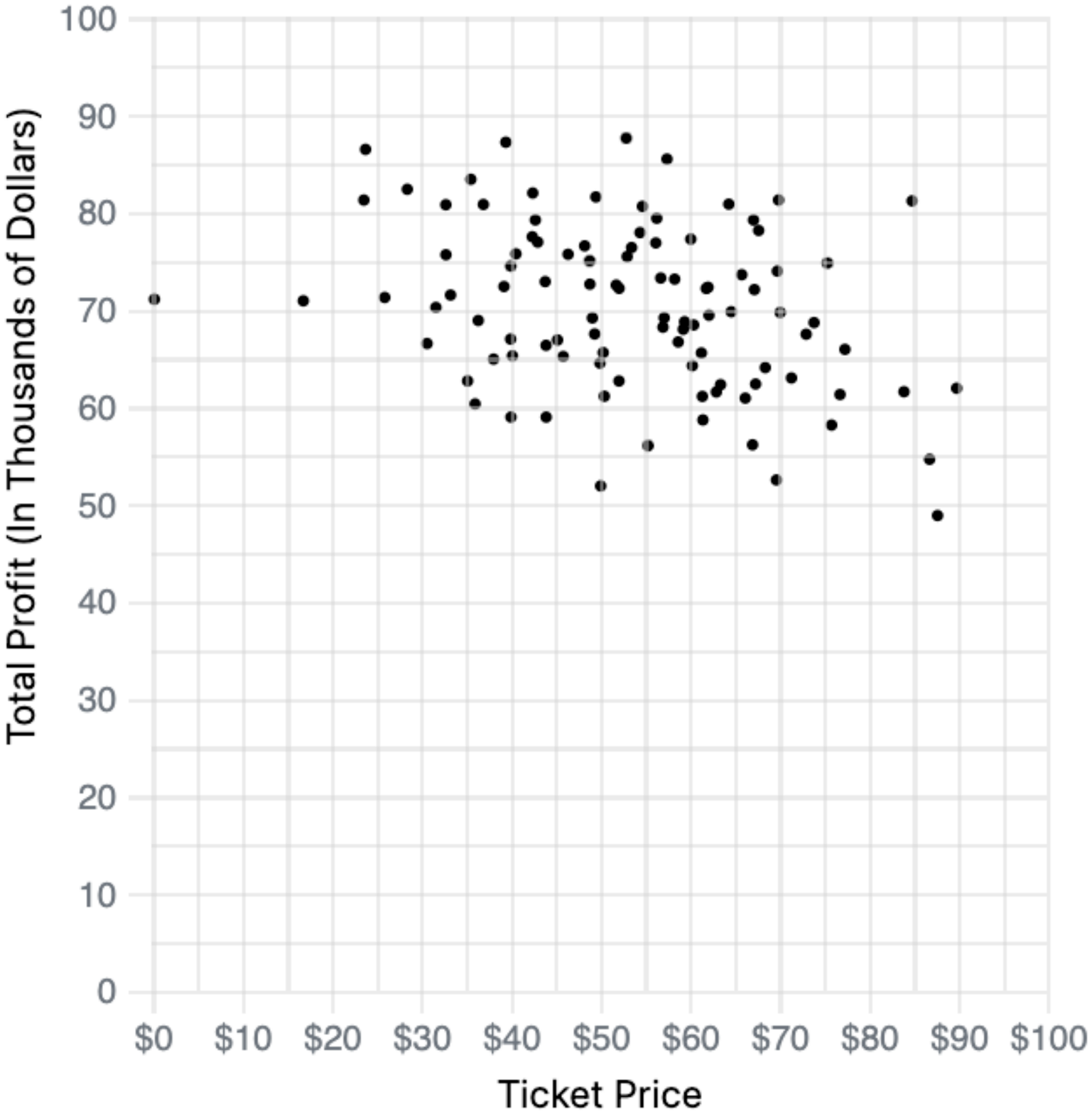}
    \caption{(left) a bad use of Grid Lines, $\alpha$=1, that is intrusive and causes the visualization to be dis-fluent, (right) a better use of Grid Lines, $\alpha$=0.45, that is not intrusive and is relatively fluent, following best practice recommendation from \cite{bartram2010whisper}.}
    \label{fig:gridlines_different_alphas}
\end{figure}

\begin{figure*}[t!]
    \centering
    \includegraphics[width = \linewidth]{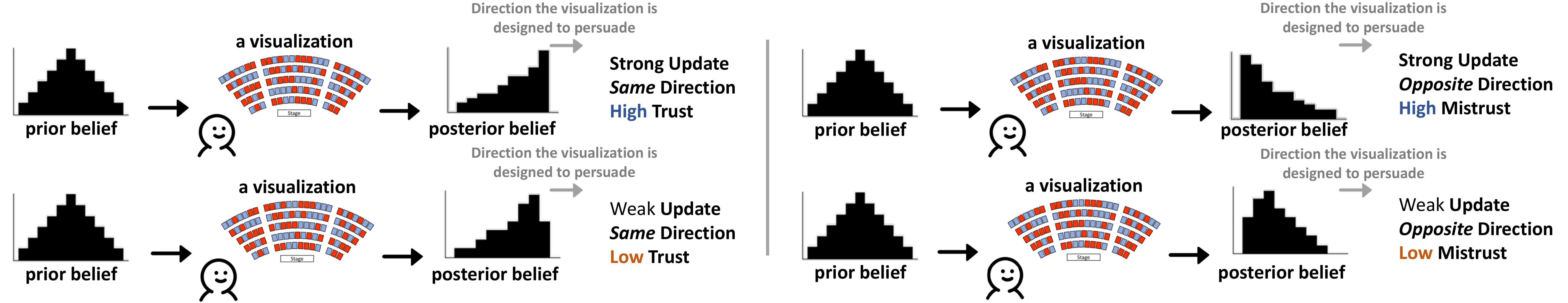}
    \caption{Examples of potential relationship between belief updating and trust.}
    \label{fig:belief}
\end{figure*}

\section{Measuring Trust via Perceptual Fluency}


Trust has been linked to the halo effect, a phenomenon coined by Edward Thorndike in which a single positive quality of a person is extrapolated to a generally positive assessment of the person in other areas \cite{thorndike1920constant}. This effect is quite significant and surprisingly is not reduced when participants are made aware of it in an experimental context \cite{wetzel1981halo}.
Beauty or aesthetics commonly produces a strong positive bias on people's impression of others. People or objects that are perceived as beautiful are also categorized as more functional and trustworthy \cite{kahneman2011thinking}. Extrapolating this effect to visualizations, we might expect beautiful visualizations would be perceived as more trustworthy.

However, a person's judgment of another's trustworthiness is formed by myriad causes, including past experiences, knowledge, culture, and even their current mood.
Psychology researchers have found a strong correlation between the perceived beauty of a subject and perceptual fluency, the ease with which a stimulus is perceived \cite{reber2004processing}.
For example, stocks with aesthetically pleasing, fluent (read: easier-to-pronounce) names and codes performed better than those with aesthetically displeasing, dis-fluent (read: hard-to-pronounce) names/codes shortly after the stocks were released to the general public when controlled for the size and industry of the companies \cite{alter2006predicting}.
Similarly, faces that are easier to categorize socially and racially are perceived as fluent and judged more attractive \cite{halberstadt2014easy}.
Hence, we would expect that more fluently-perceived people or objects are also perceived as more trustworthy.
As expected, stimuli that are more perceptually fluent (i.e., easier to perceive) are often considered more beautiful or judged more positively, leading to a positive affective association and higher trust in the stimulus \cite{graf2015dual}.
This relationship between perceptual fluency and perceived beauty or satisfaction also works in the converse direction; stimuli that are perceptually dis-fluent (visually cluttered) appear less satisfying and less trustworthy \cite{sohn2017consumer}.


Fluency also plays a role in the trust judgment of faces.
For example, Olszanowski et al. found that participants rated faces that expressed a ``pure" emotion (e.g., only one emotion, sad, happy) are easier to read, processed more fluently, and seen as more trustworthy than ``mixed" emotions (e.g., emotions that were hard to categorize as a single emotion, happy-sad) \cite{olszanowski2018mixed}. 

It is important to note that the effect of fluency on trust does not always occur. 
Oppenheimer notes people interpret fluency via naive theories regarding the cause of the fluency; in other words, people do not immediately judge fluent stimuli as fluent, but rather only after other plausible reasons for the fluency experience are not found \cite{oppenheimer2008secret}. 
For example, in the Olszanowski et al. study, when the emotions varied in dominance (e.g., angry vs. sad), fluency (i.e., ``pure" or ``mixed" emotions) did not impact the trust ratings given by participants \cite{olszanowski2018mixed}. 
Olszanowski et al. postulated that people take fluency into account for emotions that vary in valence (pleasant vs. unpleasant; attractive vs. unattractive) because fluency can be attributed to this variance (as fluency is a sort of valence metric); however, with emotions that vary in motivation, fluency is discounted as not having an effect on the level of motivation (because both sad and angry are negative valence emotions).
Hence, in an experimental setting, researchers should control for possible factors that participants might perceive as causing the fluency effect.

\subsection{How we can do this in visualization research}


More fluently processed stimuli can be perceived as more trustworthy.
This phenomenon implies that it is possible for us to measure the perceived trustworthiness of a visualization by measuring how fluently can a reader perceptually process the visualization.
Existing work has begun to show some evidence that perceptual fluency has the potential to positively influence trust in visualizations \cite{lin2021fooled}.
For example, Elhamdadi et al. measured participants' trust of fluent and dis-fluent visualizations. They found that more perceptually fluent visualizations (e.g., visualizations without heavy grid lines) are rated more trustworthy vis-a-vis trust games and subjective ratings of trust on a Likert scale \cite{elhamdadi2022trust}. 
Although it warrants further investigation, we posit that once we establish a more comprehensive model of the extent to which perceptual fluency can impact trust, we can predict how trustworthy a visualization is by measuring how fluently people perceive it.

Existing work at the intersection between data visualization and human perception has contributed concrete metrics on design factors in a visualization that can increase or decrease its processing fluency. 
For example, Bartram et al. discovered a range of alpha values for the transparency of grid lines on scatter plots that ensure the lines don't intrude on a reader's ability to interpret a visualization. 
This range of alpha values is 0.1 to 0.45, with lower alpha values preferred for sparse scatter plots and higher for dense plots \cite{bartram2010whisper}. 
Before the paper by Bartram et al., it was not uncommon to see alpha values of 1.0 as the default in visualization tools; heavy lines (with high alpha values) obscure the data and may cause the visualization to be dis-fluent. 
Because heavy grid lines create a dis-fluency effect that is unlikely to be attributed to another causal relationship, it is reasonable to expect that the dis-fluency effect will not be discounted and may cause distrust in the visualization.

As a community, we should continue to explore guidelines that can increase perceptual fluency in visualization designs and systematically test their effectiveness in improving trust in visual data communication. 
Following that, we will be able to measure the likelihood a reader will trust a visualization by proxy of perceptual fluency.

\section{Measuring Trust through Belief Updating}

In \cite{yin2019understanding}, the authors measured people's willingness to trust a machine learning model by the frequency with which people updated their predictions to match the model's output.
This approach inspires us to approximate individual trust as defined by ``the likelihood to update one's belief based on the given information.''
The more strongly someone trusts the information presented, the more likely they are to incorporate it in updating their beliefs.

The direction of the belief updating can indicate whether trust or distrust was involved.
As shown in Figure \ref{fig:belief}, if the information consumer updates their beliefs in the same direction as what the information presented, such as more firmly believing that a vaccine works after seeing information about increased vaccination rates and decreased infection rates, we can say they trust the information.
If the information consumer updates their belief in the opposite direction, such as becoming more skeptical of the vaccine after seeing the same information, we can infer that they distrust.
These results belied polarization, where two people update their beliefs in opposite directions when responding to the same evidence. Although it may seem irrational and violate the normative Bayesian, this effect occurs when interacting with controversial topics such as climate change as a product of distrust \cite{cook2016rational}.

This approach provides the potential to quantify trust. Suppose we calculate the difference between one's updated beliefs and their prior beliefs with a measure (e.g., Kullback–Leibler divergence or Earth Mover's Distance). The difference can be decomposed by the signal the person interprets from the visualization and weights (i.e., trust) that this person weighs the interpreted information to incorporate into their beliefs. While modeling these two components are empirical questions that require many controlled experiments, the lens of using belief updating to approximate trust offers the potential to further observe the degree to which the person trusts the data.



\subsection{How we can do this in visualization research}

Some visualization techniques facilitate belief-updating and, by our definition, has the potential to improve the trustworthiness of the visualization.
For example, effectively showing uncertainty in data using quantile dot plots, gradient plots, or hypothetical outcome plots can help people more appropriately update their beliefs \cite{kale2020visual, kale2018hypothetical, correll2014error,kim2020designing}. 
However, it has not yet been empirically tested whether these techniques that facilitate adequate belief-updating are also perceived as more trustworthy.
Future work should measure the ability of a visualization to update people's beliefs and the amount of trust they have in the visualization (via the other methods introduced in this paper) to model the exact relationship between belief updating and trust. 

It is also important to note that people can be biased by their knowledge and past experience when making sense of new information \cite{dieckmann2017seeing, kahneman2011thinking}, such as exhibiting confirmation bias to overweigh information that is congruent with their prior beliefs and under weigh information that is incongruent \cite{klayman1995varieties}.
Therefore when attempting to measure trust via measuring belief updating, it is critical for us to consider valid and reliable methods of belief elicitation. 
We can capture people's beliefs by asking people to manipulate the slope of a line in a chart ~\cite{karduni2020bayesian, kim2017explaining, mantri2022how}, indicating the range of their belief on a Likert scale \cite{kim2020bayesian}, or to make predictions \cite{griffiths2006optimal}.
Future work can compare these channels of belief measurements with each other to refine metrics that directly map with trustworthiness.

Existing work has demonstrated that people can update their belief on the relationship between two entities after seeing just one instance of a visualization \cite{xiong2022seeing}.
So while the amount of change between prior and posterior beliefs can likely be used as a proxy to determine how much people trust the new data they saw, we should also consider the possibility of people updating their beliefs due to mere exposure effects ~\cite{harrison1977mere}, and account for them when we create quantitative models of belief updating and trust.


\section{Conclusion}


Trust is a complex topic.
How trust impacts human-data interactions (and by extension, the trust relationship between the reader and the visualization designer) remains mostly under-explored. 
The visualization community does not yet have a systematic understanding of what factors impact trust in visual data communication nor a formalized model for establishing trust between humans and data. 

In this paper, we presented a multidisciplinary, but by no means exhaustive, collection of trust measurements from existing work in computer science, psychology, behavioral economics, and other social sciences.
We call for visualization researchers to consider these methods, adapt them for visualization research, and, as a community, iteratively improve them for reliability and validity.
This paper is our first step in an ongoing process toward defining and modeling trust in visual data communication, the future result of which we hope will be a set of design guidelines for enhancing trust.

With these trust measurements, we can develop a better understanding of trust in human-data interactions.
Establishing this understanding can further inform visualization practices in many sub-fields of computer science. 
For example, visualization is increasingly used in the field of explainable artificial intelligence to describe the inner workings of machine learning algorithms, and efforts such as the TRust and EXpertise in Visual Analytics (TREX) Workshop \cite{trexvis} continue to promote interdisciplinary science to support human-centered AI research and practice. 
Presenting data visualizations in a trustworthy manner can enhance understanding and engagement, preventing potential confusion or misuse of the algorithms and AI tools.


\bibliographystyle{abbrv-doi-hyperref-narrow}

\balance

\bibliography{template}
\end{document}